\documentclass[aps,preprint,prd,epsfig]{revtex4}
\usepackage{amssymb}
\usepackage{amsmath}
\usepackage{color}
\usepackage{fancyhdr}

\usepackage{slashed}
\usepackage{graphicx}
\usepackage{epsfig}
\usepackage{amssymb}
\usepackage{bm}
\usepackage{color}
\usepackage[dvipsnames]{xcolor}
\usepackage{hyperref}
\def\br{\begin{eqnarray}}
\def\er{\end{eqnarray}}
\def\be{\begin{equation}}
\def\ee{\end{equation}}

\def\({\left(}
\def\){\right)}
\def\<{\left\langle}
\def\>{\right\rangle}

\usepackage{slashed}
\usepackage{amsmath}
\usepackage{subfigure}
\usepackage{graphicx}
\usepackage{epsfig}
\usepackage{amssymb}
\usepackage{dcolumn}
\usepackage{bm}
\usepackage{color}

\newcommand{\ba}{\begin{array}}
\newcommand{\ea}{\end{array}}
\def\br{\begin{eqnarray}}
\def\er{\end{eqnarray}}
\def\be{\begin{equation}}
\def\ee{\end{equation}}

\def\({\left(}
\def\){\right)}

\def\<{\left\langle}
\def\>{\right\rangle}

\def\tt{\textnormal\tiny\textsc}

\begin{document}

%\markboth{A. Doff and A. A. Natale}
%%%%%%%%%%%%%%%%%%%%% Publisher's Area please ignore %%%%%%%%%%%%%%%
%
%\catchline{}{}{}{}{}
%
%%%%%%%%%%%%%%%%%%%%%%%%%%%%%%%%%%%%%%%%%%%%%%%%%%%%%%%%%%%%%%%%%%%%
\title{Composite scalar boson mass dependence on the constituent mass anomalous dimension }

\author{A. Doff$^{a}$ and  A. A. Natale$^{b}$}
%\email{cpires@fisica.ufpb.br}

\affiliation{$^a$Universidade Tecnol\'ogica Federal do Paran\'a - UTFPR - DAFIS, R. Doutor Washington Subtil Chueire, 330 - Jardim Carvalho,  84017-220
Ponta Grossa, PR, Brazil\\ 
agomes@utfpr.edu.br \\
$^b$ Instituto de F{\'i}sica Te\'orica - UNESP, Rua Dr. Bento T. Ferraz, 271,\\ Bloco II, 01140-070, S\~ao Paulo, SP, Brazil\\
natale@ift.unesp.br }

%\date{\today}

%\begin{history}
%\received{Day Month Year}
%\revised{Day Month Year}
%\end{history}

\begin{abstract}
We perform a Bethe-Salpeter equation (BSE) evaluation of composite scalar boson
masses in order to verify how these masses can be smaller than the composition scale. The
calculation is developed with a constituent self-energy dependent on its mass anomalous
dimension ($\gamma$), and we obtain a relation showing how the scalar mass decreases as $\gamma$ is
increased. We also discuss how fermionic corrections to the BSE kernel shall decrease the
scalar mass, whose effect can be as important as the one of a large $\gamma$. An estimate
of the top quark loop effect that must appear in the BSE calculation gives  a lower bound
 on the composite scalar mass. 
%\keywords{Bethe-Salpeter equation; composite scalar bosons;  nonperturbative techniques; nonperturbative calculations.}
\end{abstract}
\maketitle

%\ccode{PACS numbers:12.60.Nz,12.60.Cn, 12.60.Rc,11.15.Tk.}

%%%%%%%%%%%%%%%%%%%%%%%%%%%%%%%%%%%%%%%%%%%%%%%%%%%%%%%%%%%%%%%%%%%%%%%%%%%%%%%%%%%%%%%%%%%%%%%%%%%%%%%%%%%%%%%%%%%%%%%%%%%%%%%%%%%%%%%

\section{Introduction}

The discovery of the Higgs boson at the LHC \cite{atlas,cms} completed the Standard Model (SM), where a scalar boson sector is present as proposed long ago \cite{sw}. In many extensions of this model this scalar sector is even larger than the SM one, although experimental signals of new particles belonging to this sector
are still missing. Furthermore, there are theoretical shortcomings about this scalar sector \cite{kw,gh}.

The absence of signals of a large scalar boson sector in the experimental data, as well as a possible explanation of a light Higgs boson has been discussed recently in Ref.\cite{el,lane}. Composite scalar bosons also appear in the context of Technicolor theories (TC) \cite{wei,sus,far}, which usually have a composition scale of order of $\Lambda_H \geq 1$TeV. 

The construction of a realistic Technicolor model may indeed be a very precise engineering problem. In order to get around this problem different models with large mass anomalous dimension ($\gamma$) emerged, as walking technicolor\cite{lane0,appel,aoki,appelquist,shro,kura}  and gauged Nambu-Jona-Lasinio models\cite{yama1,yama2,yama22,yama3,mira3,yama4,takeuchi}. A light composite scalar boson may be generated when the strong interaction theory (or TC) has large mass anomalous dimension ($\gamma$) as proposed by Holdom\cite{holdom}. A discussion of how these scenarios can lead a light composite scalar boson became clear in the Refs.\cite{rev1}. Based on these scenarios, calculations involving effective Higgs Lagrangians have led to different predictions regarding to composite scalar bosons masses.

In this case the self-energy of the new fermions (or technifermions) responsible for the composite states is characterized by a large mass anomalous dimension,
resulting in mass diagrams whose calculation do not scale with the naive dimensions. This self-energy at large momenta is proportional to
\be
\Sigma_{\tt{C}} (p^2) \propto \frac{\mu_{\tt{C}}^3}{p^2} (p^2/\mu_{\tt{C}}^2)^{\gamma/2} .
\label{eq0}
\ee
where $\mu_{\tt{C}}$ is the typical composition or dynamical mass scale and $\gamma$ the mass anomalous dimension.
It is now known that according to the mass anomalous dimension, this self-energy may vary asymptotically from a $1/p^2$ behavior, up to a very slowly decreasing
logarithmic behavior with the momentum \cite{takeuchi,us}. It is usually assumed that large $\gamma$ values appear in gauge theories with higher fermionic
representations or with a large number of fermions. However, Eq.(\ref{eq0}) can be quite modified just coupling two strongly interacting theories \cite{us,us1}.

A light composite scalar boson appears naturally when its mass is calculated with the help of Eq.(\ref{eq0}) (see, for instance, Ref.\cite{rev1}). It is interesting to  see that the problem of understanding a possible light composite Higgs boson could also be related to the case of the sigma meson.
The sigma meson is the QCD scalar composite now known as $f_0(500)$. A standard Bethe-Salpeter equation (BSE) calculation gives $m_\sigma = 670$MeV~\cite{blat}, which is larger than its experimental value. The detailed work of Ref.\cite{blat} deals with possible contributions that may lower the estimate of this mass.
A composite $J=0$ state may have many contributions to its mass. In the case of the sigma meson it is not even clear how much of its composition is due to different
quarks, even more the amount of its mass that is due to gluons, although it is already a puzzle the fact that a simple BSE mass estimate gives a result larger
than the experimental value. Therefore, it is natural to think what a calculation similar to the one of Ref.\cite{blat} would teach us about the composite Higgs mass. 

As far as we know there are not in the literature detailed calculations of the composite Higgs boson mass using BSE, particularly looking for effects of different
$\gamma$ values or the effect of massive fermions contribution to the BSE kernel. In this work we calculate the scalar composite mass with the help of Bethe-Salpeter equations, obtaining a relation between the scalar mass and $\gamma$. We present a fit of such relation expecting that it could be tested by other methods. This calculation is shown in Section II. In Section III we perform a simple order of magnitude estimate of fermionic loop effects that contribute to the BSE kernel and can lower the scalar mass value. It is important to remark that most studies dealing with the possibility of a light composite scalar boson are based on the calculation of Schwinger-Dyson equations and related to a near conformal behavior of the theory \cite{hold2}. The advantage of the BSE approach is that it takes into account all the possible bound states that contribute to the scalar mass calculation. A light scalar comes out not only due to a conformal behavior, but it does emerge when some bound states contribute negatively to the scalar BSE kernel. We verify that the effect of a large fermionic mass, particularly a bound state formed by the top quark, affect the Higgs boson mass calculation and cannot be neglected. A quite simple estimate of the top quark contribution to the BSE kernel leads to a lower bound on the composite Higgs mass of the order of 123.3GeV. As far as we know there are not in the literature detailed calculations of the composite Higgs boson mass using BSE. In this work we generalize the results obtained in \cite{us3}, improving the work of Ref.\cite{jm2}, which is applied only to QCD bound states, introducing fermionic propagators characterized by Eq.(\ref{eq0}) which is dependent on different anomalous mass dimension, and investigating the effect of TC bound states in order to obtain a light composite scalar mass. Section IV contains our conclusions.

\section{The scalar mass: BSE and mass anomalous dimension}

The scalar boson mass can be calculated using BSE once we specify the propagators and vertices of the strong interaction that binds such scalar boson.
For the scalar case the Bethe-Salpeter equation has the form\cite{jm1,jm2,jm3}
\br
&&\chi (p,q) = -\imath \int \frac{d^4k}{(2\pi)^4} S(q+\alpha p)K_{\rho\nu}(p,k,q) S(q-\beta p) ,\nonumber \\
&& K_{\rho\nu}(p,k,q) = \gamma_\rho \chi (p,k) \gamma_\nu G_{\rho\nu} (k-q)  ,
\label{eq1}
\er
where $\alpha + \beta =1$ ($\alpha$ and $\beta$ characterize the fraction of momentum carried by the constituents), although the result is not dependent
on these quantities\cite{jm2}. As a first approximation we shall choose $\alpha = \beta = 1/2$.
$G_{\rho\nu}$ is the gauge boson propagator in the Landau gauge. There are different models for this quantity which will be discussed ahead. 

The fermion propagator is given by
\be
S^{-1}(q) = {\slashed{q}} A(q^2) - B(q^2) \, ,
\label{eq2}
\ee
From now on we shall assume $A(q^2)=1$ and $B(q^2)=\Sigma (q^2)$. Usually in this type of calculation the self-energy $\Sigma (q^2)$ is obtained from the
numerical solution of the Schwinger-Dyson equation (SDE) for the fermionic propagator. In order to simplify the calculation this self-energy is going to be given by one 
ansatz that is a function of the mass anomalous dimension $\gamma$.

We recall that $\Sigma (q^2)$ at large momenta in QCD (or any asymptotically free non-Abelian gauge theory) is given by \cite{politzer}
\be
\Sigma (q^2) \propto \frac{\<{\bar\psi}\psi\>_\mu}{q^2} \(\frac{q^2}{\mu^2}\)^{\gamma/2} \, ,
\label{eq3}
\ee
assuming for the fermionic condensate $\<{\bar\psi}\psi\>_\mu \approx \mu^3$ where $\mu$ is the dynamically generated mass.

The ansatz that we assume for the self-energy has the form
\be
\Sigma (q^2) = \frac{\mu^3}{q^2+\mu^2} \( \frac{q^2+\mu^2}{\mu^2}\)^{\kappa} \, 
\label{eq4}
\ee
where 
\be
\kappa = \gamma /2.
\label{eq4l}
\ee 
Eq.(\ref{eq4}) behaves in the infrared region as a mass $\mu$ and decays at large momenta as prescribed in Eq.(\ref{eq3}), maping the SDE in
the full Euclidean space, and will allow us to obtain BSE solutions for the scalar mass as a function of $\gamma$. Note that 
$\kappa$ is just a parameter that when $\kappa =0$ the self-energy behaves asymptotically as $1/p^2$, and when $\kappa \rightarrow 1$ Eq.(\ref{eq4}) 
behaves like
\be
\Sigma (q^2)\approx \mu \left[ 1+ \delta_1\ln\left[(q^2+\mu^2)/\mu^2 \right] \right]^{-\delta_2} \, ,
\label{eq5}
\ee
where $\delta_1$ and $\delta_2$ are obtained from $\gamma$ when expanded as a function of the running coupling $g^2(q^2)$.

The BSE solution appear as an eigenvalue problem for $p^2=M^2$, where $M$ is the bound state mass. The variables are $p,q,k$. $k$ is integrated
and we remain with a equation in $q$ that will have a solution for $p^2=M^2$. In the Eq.(\ref{eq1}) $\chi$ can be projected into four
coupled homogeneous integral equations, that  implies for projection of the scalar component 
\be
\chi (p,q)=\chi_{S0}+\slashed{p}\chi_{S1}+\slashed{q}\chi_{S2}+
[\slashed{p},\slashed{q}]\chi_{S3} \, ,
\label{eq6}
\ee 
which are functions of $p^2$, $q^2$ and $p.q=pqcos\theta$.

\par It is possible to expand $\chi (p,q)$ in terms of  Tschebyshev polynomials,  and  these equations can be truncated at a given
order determined by the relative size of the next-order functions. In accordance with Ref.\cite{jm1}, a satisfactory solution can be obtained
by keeping only some terms, like $\chi_{S(0,1)}^{(0)},  \chi_{S(1)}^{(0,1)},  \chi_{S(2)}^{(0,1)}$. 

To set up the problem we follow closely the work of Ref.~\cite{jm2}, where the BSE was solved for the scalar boson constituents with masses
$m_a$ and $m_b$, and for simplicity we assume
\be
m_a = m_b = m= \Sigma (x+\frac{1}{4}p^2) ,
\label{eq7}
\ee
where it was assumed $\alpha = \beta$ (each constituent carries half of the momentum), and $x=q^2/\Lambda^2$, where $\Lambda = \Lambda_{QCD}$ (or $\Lambda =
\Lambda_H$, the TC mass scale) the characteristic mass scale of the bounding force. 

The different components of Eq.(\ref{eq6}), $\chi_{S(i)}^{(0,1)}$ for $i=0..3$,  are given by 
\be
\chi^{(0)}_{S0}=2[(x-\frac{1}{4}p^2-m^2) J_1] I_{S0} + 
\Delta\chi^{(0)}_{S0} \, ,
\label{eq9}
\ee
with
\be
I_{S0} = \frac{2}{3\pi}\int dy y \chi^{(0)}_{S0} K_1 \, ,
\label{eq10}
\ee
where $y=k^2/\Lambda^2$ and 
\be
K_1(x,y)=\frac{3}{16\pi^2} \int d\theta sen^2 \theta G(x,y,cos \theta) \, ,
\label{eq11}
\ee
\be
J_1= \frac{2}{\pi} \int_0^\pi d\theta\frac{sen^2 \theta}{D(p^2,q^2, pq cos\theta )}
\label{eq12}
\ee
 and in the Eq.(\ref{eq12})
\be
D(p^2,q^2, pq cos\theta )=\{ (q+\frac{1}{2}p)^2+m^2 \}\{ (q-\frac{1}{2}p)^2+m^2 \}.
\label{eq8a}
\ee
\par In the above equation we can expand $(q+\frac{1}{2}p)^2+m^2$ and $(q-\frac{1}{2}p)^2+m^2$  in  Taylor series.
Just keeping the first-order  derivative terms for $m$, we have that the function $J_1$ can be written in the form

\br
J_1 = \frac{2}{c_1c_4 + c_2c_3}\left[\frac{c_2}{D_1}+ \frac{c_4}{D_2} + d_1\left(\frac{c_1}{D_1} -\frac{c_3}{D_2}\right)\right]
\er
where, in our approximation, we have
$$ c_1=c_3 = x+\frac{1}{4} p^2 + m^2$$
$$ c_2=c_4 = 1+2m m^\prime$$
and $m^\prime$ is the derivative of $m$ with respect to the momentum. In addition, as a consequence of $\alpha=\beta$ we obtain
$$ d_1=0$$
$$ D_1=D_2= c_1+\sqrt{c_1^2-p^2xc_2^2}. $$

\par In Eq.(\ref{eq9}) , the term $\Delta\chi^{(0)}_{S0}$  represent corrections to the leading-order results of $\chi^{(0)}_{S0}$,
that correspond to  $\chi^{(0,1)}_{S1}$ , $\chi^{(0,1)}_{S2}$ and $\chi^{(0,1)}_{S3}$. With the approximations considered here we obtain  

\be
\begin{split}
\Delta\chi^{(0)}_{S0} = -\frac{4}{3\pi} m J_1 \int dy y \chi_{S2}^{0} \sqrt{xy} (3K_6 -2\sqrt{xy}K_3) - \\
-\frac{2}{3\pi} p^2 (J_3 - J_1) \int dy y \chi_{S3}^{0}\{ 2\sqrt{xy} K_6 - \frac{8}{3} xy K_3 \} - \\
-\frac{4}{\pi}\{ (-x+\frac{1}{4}p^2+m^2)J_2 \} \int dy y \chi_{S0}^{(1)}\sqrt{\frac{y}{x}} K_6 - \\
-\frac{2}{\pi}\{ (-x+\frac{1}{4}p^2+m^2)J_1 \} \int dy y^2 \chi_{S0}^{(2)} \{ \frac{4}{3} K_7 - K_1 \}.
\end{split}
\label{eq13}
\ee

\par In the Eq.(\ref{eq13}), the lowest order terms $\chi_{S(1-3)}^{(0)} $ are given by 
\be 
\chi_{S1}^{(0)}= 0 
\ee 
\br
&&\chi_{S2}^{(0)}=-\frac{4}{\pi} m J_1 I_{S0}+
\frac{2}{3\pi}\(-x-\frac{1}{4}p^2+m^2\)\times \nonumber \\
&&\,\,\,\,\,\,\,\,\,\,\,\,\int dy y \chi_{S2}^{0} \( 3\sqrt{\frac{y}{x}}K_6 -2y K_3 \),
\label{eq14}
\er
 and 
\be
\chi_{S3}^{(0)}= \frac{3}{2}J_1 I_{S0}. 
\label{eq15}
\ee
While the higher order terms are described by
\be
\chi_{S0}^{(1)} = -6 \( (-x+\frac{1}{4}p^2+m^2)\frac{J_2}{xp^2} \) I_{S0} ,
\label{eq16}
\ee
and 
\be
\chi_{S0}^{(2)}=-3\{\frac{1}{xp^2}\(-x+\frac{1}{4}p^2+m^2\)(4J_3-J_1)\} I_{S0}.
\label{eq17}
\ee
\par  Since we are dealing with a scalar boson case with equal mass constituents the equations of Ref.\cite{jm2} are also
simplified and we have:
\be
J_2 = 0
\label{eq18}
\ee

\be
J_3= \frac{1}{D_1^2}
\label{eq19}
\ee
\noindent where in the Eqs.(\ref{eq13}-\ref{eq14})

\be
K_3(x,y)=\frac{3}{16\pi^2} \int d\theta \frac{sen^4 \theta}{x+y-2\sqrt{xy} cos\theta} G(x,y,cos\theta) ,
\label{eq20}
\ee

\be
K_6(x,y) = \frac{3}{16\pi^2} \int d\theta sen^2 \theta cos\theta G(x,y,cos\theta) ,
\label{eq21}
\ee

\be
K_7(x,y) = \frac{3}{16\pi^2} \int d\theta sen^4\theta G(x,y,cos\theta) .
\label{eq22}
\ee

These equations can be solved making use of suitable expressions for the main Green functions (i.e. propagators and vertices). The gauge
boson propagator is given by
\be
G(k^2)= \frac{16\pi}{3} \left[\frac{\pi d}{k^2 \ln(x_0+x)} \right] +  G_{IR}(k^2),
\label{eq23}
\ee

\noindent  where $G(x,y,cos\theta) = \Lambda^2 G(k-q)^2$, and  $ G_{IR}(k^2)$ is an assumed form of the interaction at infrared momenta. In Ref.\cite{jm2} this 
contribution was chosen to be of the form
\be 
 G_{IR}(k^2) = \frac{16\pi}{3}ak^2e^{-\frac{k^2}{\omega^2}},
\label{eq23b}
\ee 
and in the  gaussian ansatz for the $ G_{IR}(k^2)$ , $\omega \in [0.4, 0.6] GeV$\cite{RP,RP2,RP3}.  According to Ref.\cite{jm2}
the parameters used in the QCD case are given by 
\br 
&& a=(0.387 \, GeV)^{-4} \,\,\,,\,\,\,\,\, \omega= (0.510 \, GeV) \nonumber \\
&& d= 12/(33-2n_f) \,\,\,\,\,\,,\,\,\,\,\, n_f=5 \nonumber\\
&& \mu \approx  \Lambda_{QCD} = 0.228 GeV  \,\,\,\,\,\,,\,\,\,\,\,  x_0 = 10.  
\er 

\par  More recent expressions for these Green functions were formulated in Refs.\cite{RP,RP2,RP3} , where the two free parameters in Eq.(\ref{eq23b}), $\omega$ and $a=D$,  are  parameterized by $(\varsigma_g)^3 = D\omega= const$, and  the fitted values of $\varsigma_g$ depend on the form that is assumed for the dressed-gluon quark vertex. Considering the recent results reported in Ref.\cite{RP}, it is possible to verify that these choices do not modify substantially the numerical results.

\par The complete set of BS equations given by Eq.(\ref{eq1}) was solved numerically by iteration starting with the inhomogeneous terms as input, where the
function $\chi(p, 0) = \chi^{(0)}_{S0}$ is fixed to some arbitrarily chosen value,  considering an interactive process in the equation below
\br
\Delta(p^2) = \chi (p^2,0) - \chi (p^2,q^2)
\label{eqD}
\er 
\noindent  to find the eigenvalues $p^2=M^2_{S}$, of eigenfunction $\chi (p^2,q^2)$. For a given value of $p^2$ , not an eigenvalue, $\Delta(p^2)$  is not zero, then after several interactions we can finding $p^2$ such that we obtain $\Delta(p^2) = 0$, or BS wave function eigenvalues. Notice that a normalization procedure is also necessary because we cannot have arbitrary mass anomalous dimensions, as already pointed out in Refs.\cite{us3,lane3}. 
 
Assuming Eqs.(\ref{eq6} - \ref{eq23b}) , in Fig.1 we present the results obtained for $\Delta(p^2)$, given by Eq.(\ref{eqD}), considering the anzats 
Eq.(\ref{eq4}) for $m$. In this figure we normalize ours results for $M_S$ in terms of
$$
M_S = 2\Lambda_{QCD},
$$
associated to a negligible $\gamma$.
\begin{figure}[t]
\begin{center}
\includegraphics[width=0.6\columnwidth]{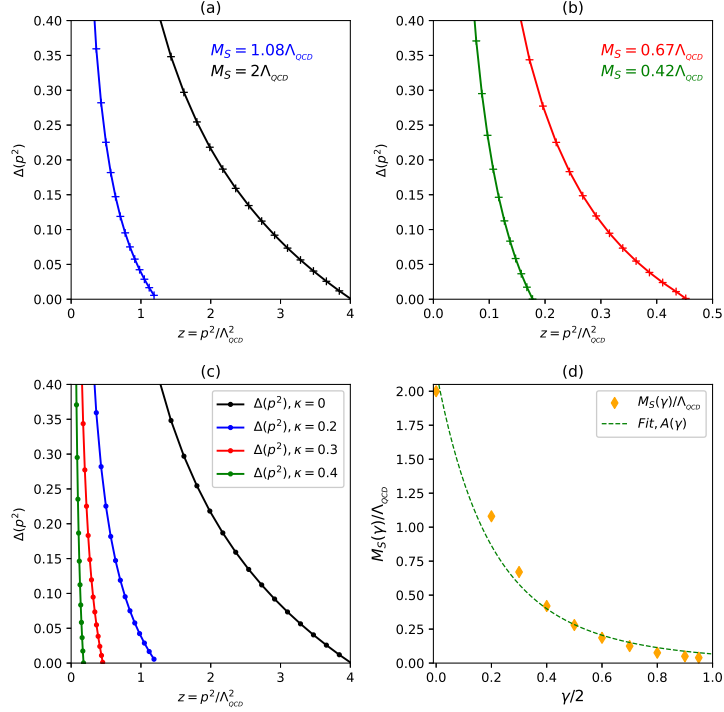}
\caption{In this figure  we show the scalar masses $M_S(\gamma)$ obtained for $\Delta(p^2)$, given by Eq.(\ref{eqD}), in the QCD case. The contextualization of the curves behavior are described in the text. }
\end{center}
\end{figure}
\par The choice of this normalization is based on the result described in Ref.\cite{DS}, where Delbourgo and Scadron  verified analytically with the
help of the homogeneous Bethe-Salpeter equation (BSE) , that the sigma meson  mass is given by  $m_{\sigma} = 2\mu_{dyn}$. In this calculation it is assumed
that the dynamically generated quark mass behaves (for large $p^2$) as  $m_{dyn}\sim \frac{\mu^3_{QCD)}}{q^2}$, which corresponds to the case where $\kappa=0$ in the ansatz proposed in Eq.(\ref{eq4}). In this way, based on this normalization, we can follow the behavior of how  $M_S$ resulting from Eq.(\ref{eqD}) is influenced by $\gamma$.

\par  In Fig.(1a) the black line corresponds exactly to the case where $\kappa=0$, while the blue line to $\kappa=0.2$ . In Fig. (1b) we consider the cases where $\kappa=0.3$(red line) and $\kappa =0.4$(green line), the  Fig.(1c) is a composition of the previous results, where we indicate for each curve the value assumed for $\kappa$. Note that in Fig1.(a-b) , in the upper right corner, we describe $M_S$ found in each case observing that for a given $z = p^2/\Lambda^2_{QCD} $ we have 
\be 
M_S = \sqrt{z}\Lambda_{QCD},
\ee 

\noindent Finally, in Fig.(1d), we present the behavior of $M_S(\gamma)$ obtained for $\kappa$ in the range $\in [0,0.95] $. We obtained a very simple fit
to the data 
with $R^2 = 0.977$ which corresponds to
\be
A(\gamma) = \frac{M_S(\gamma)}{\Lambda_{QCD}} = \frac{2.15}{(1 + \gamma/2)^{5.34}}.
\label{eqg1}
\ee 
It is clear that Fig.(1d) may be slightly dependent on the propagators and vertex that we assumed here. However, we expect that the behavior of this curve
can be tested by other methods, and more importantly it shows how the scalar composite mass should behave as we vary the constituent mass anomalous dimension.

\begin{figure}[t]
\begin{center}
\includegraphics[width=0.6\columnwidth]{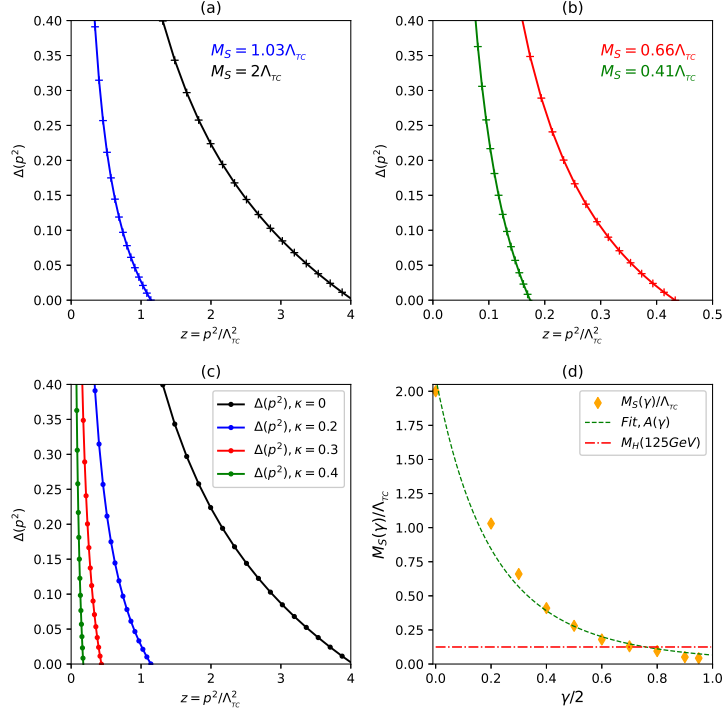}
\caption{In this figure  we show the scalar masses $M^{TC}_S(\gamma)$ obtained for $\Delta(p^2)$, given by Eq.(\ref{eqD}), in the TC case. The contextualization of the curves behavior are described in the text.}
\end{center}
\end{figure}

\par We can now focus on the Higgs boson case. In Fig.2 we extend the results obtained  for QCD in the case of a  $SU(3)$ TC model. As a first approach, since TC is based upon an analogy with the dynamics of QCD, we can use the equations obtained for QCD to determine, by appropriate rescaling, the behavior of $M^{TC}_S(\gamma)$. 
Hence, we can estimate  $\omega_{TC}$, $a_{TC}$ from the QCD analogue using the following scaling relation
\be 
(\omega , a )_{TC} = \sqrt{\frac{N_{TC}}{3}}\frac{{\Lambda_{TC}}}{\Lambda_{QCD}}(\omega, a)_{QCD}, \nonumber \\
\ee

\noindent where  $\Lambda_H = \Lambda_{TC}= 1TeV$. In this case the results for $M_S(\gamma) = M^{TC}_S(\gamma)$ follow from the normalization $M_S = 2\Lambda_{TC}$; and we include in Fig.(2d) the dot-dashed line in red, which corresponds to the observed Higgs boson mass for the purpose of comparison with the $M_S(\gamma)$ behavior. In the region where $\kappa \sim 0.8$, we recover the result obtained for the extreme walking behavior\cite{us3},  where for a $SU(3)$ TC model in that reference,  assuming for $\Sigma(p^2)_{TC}$ the behavior given by Eq.(\ref{eq5}), we obtained $M_{H} \sim O(110) GeV$.

\par  In Fig.(2d), we present the behavior for $M^{TC}_S(\gamma)$ obtained in the range $\kappa \in [0,0.95] $. The fit obtained with $R^2 = 0.988$ corresponds to
\be
A(\gamma) = \frac{M^{TC}_S(\gamma)}{\Lambda_{TC}} = \frac{2.1}{(1 + \gamma/2)^{5.12}}.
\label{eqg2}
\ee 
This result shows how the composite Higgs boson mass may vary with the constituent mass anomalous dimension. However, as we will discuss in the next section, it is not only the value of $\gamma$ that modifies these estimates. Note that Eq.(\ref{eqg2}) do not differ appreciably from Eq.(\ref{eqg1}) due to the fact that
we have chosen the TC gauge theory as a QCD rescaled version with self-energy given by ansatz, Eq.(\ref{eq4}). 

Notice that $\Delta(p^2)$  and the scalar masses in Fig. 1 and Fig. 2 decay faster with increasing gamma. The fact is that the BSE kernel
is proportional to the fermionic and gauge boson propagators, and is an integration over these quantities as well over the coupling constants. The product
of coupling and gauge boson propagator may be interpreted as the strength of the interaction, and the effect of larger $(\gamma)$ values in the fermionic propagators
act in the sense of diminish the interaction strength implying in smaller scalar masses, and similarly a change in the composite scalar wave functions.

\section{Scalar masses: the effect of fermions}

In usual BSE calculations of QCD light hadronic states the effect of heavy quarks are not included. However, when performing a BSE calculation of a possible
composite Higgs boson we cannot neglect the contribution of a top quark loop to the BSE kernel. There are two reasons for this; the strong coupling of the top
quark to the scalar boson and the approximate values of the masses of these two particles. We are clearly interested in the case of a composite Higgs
boson mass calculation, but we shall start with a simple discussion about the QCD scalar (the sigma meson), for which a more detailed calculation will
be left for a forthcoming work.

If we go back to the linear sigma model at 
constituent level we know that the sigma couples to fermions as 
\be
{\cal{L}_{\sigma}} = \lambda \bar{\Psi}\Psi\phi ,
\label{eq24}
\ee
where $\phi$ stands for $\sigma$. This coupling imply that the $\phi$ mass obtains contributions from the BSE diagram shown in Fig.(3a), that comes with
a negative sign due to the effect of a fermion loop. Eq.(\ref{eq24}) also describe the Higgs Yukawa coupling to fermions. In particular, when we consider a
composite Higgs boson the coupling to fermions is more sophisticated, we may even have fermionic contributions to the BSE like the one shown in
Fig.(3b), involving the exchange of extended TC gauge bosons (ETC)~\cite{far}. However, as the gauge bosons of Fig.(3b) are very heavy, the vertex in that figure can be reduced to an effective vertex as shown in Fig.(3c) and
the final BSE mass contribution can be reduced to the one of Fig.(3a).

The fermionic contribution to the BSE would be given by
\br
&&\Pi (p^2) = m_{f(t)}^2 Tr \int_0^{\Lambda} d^4 q \chi(q^2){\cal F}(q,p,f(t)) \chi (q^2) , \nonumber \\
&& {\cal F}(q,p,f(t))=  S_{f(t)}(p/2+q) \chi (q^2) S_{f(t)}(p/2-q)
\er
where the vertex of the BSE (due to a fermion $f$ or to the top quark $t$) reads
\be
m_{f(t)} \times \chi(p^2) ,
\ee
where we stress the effect of the large top quark mass in the calculation of the Higgs boson mass. A full calculation of the BSE including fermionic corrections with
complete solutions of the Schwinger-Dyson equations for the self-energies is a lengthy work and is under study; it may affect the sigma as well as the Higgs boson mass estimate. In the case of the Higgs boson we can resort to a simple estimate of the loop of Fig.(3a), i.e. a correction of $M^2_{\phi}$ represented by $\delta M^2_{\phi}$, is given by

\begin{figure}[t]
\centering
\includegraphics[width=0.45\columnwidth]{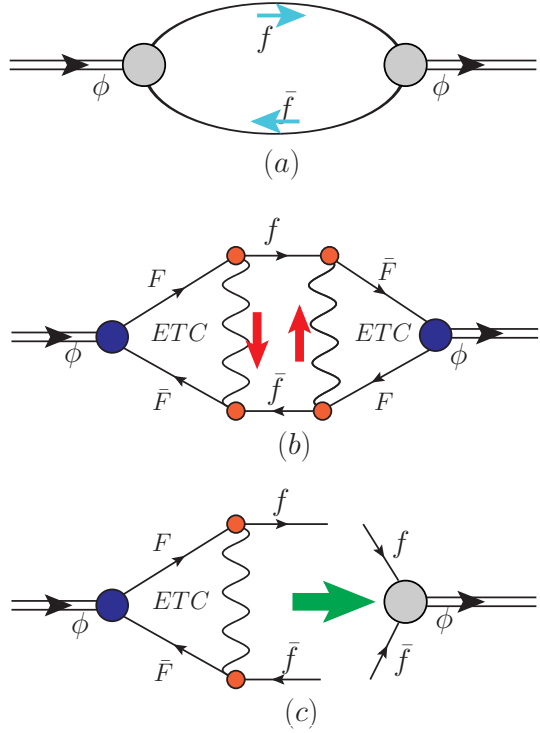}
\caption[dummy0]{In Fig.(3b) we indicate the BSE diagram that should introduce fermionic corrections to $M^2_{\phi}$ in the case of a composite Higgs boson. As the ETC gauge bosons depicted in this diagram are very heavy, the vertex appearing in Fig.(3b) can be reduced to the one of Fig.(3c). Therefore, Fig.(3a) is the result of
Fig.(3b) when we assume the effective vertex, and it reduces the scalar boson mass.}
\label{figmass}
\end{figure}

\be
i\delta M^2_{\phi} = \frac{\lambda^2 N_f}{(2\pi)^4}\int d^4q \frac{\Sigma^2(q^2)}{( q^2-\Sigma^2(q^2))^2} ,
\label{eq25l} 
\ee
where $N_f$ is the number of fermions (f) in the loop, and the  biggest  effective  coupling $\lambda$ (when f= top quark)  is given by
\be
\lambda = m_t (\gamma) / v ,
\label{eqmt}
\ee
and $v$ is the standard model vacuum expectation value ($v = 2M_W/g_w$).

Eq.(\ref{eq25l}) is enough to verify the order of magnitude that we shall obtain when solving the complete system of BSE for the scalar boson. Considering the anzats described by Eq.(\ref{eq4}), in euclidean space we obtain the following expression for $\delta M^2_{\phi}(\gamma)$
\br 
\delta M^2_{\phi}(\gamma) = \frac{\lambda^2 N_f}{4\pi^2} f(\gamma) \Lambda^2_{TC}
\label{eq29}
\er 
\noindent where 
\be 
f(\gamma)  = \frac{1-\gamma}{3 - \gamma}. 
\label{eq29a}
\ee 
The behavior of Eq.(\ref{eq29a}) with $\gamma$ is an artifact of our approximation, since the $\gamma$ running with
momentum was not considered, and its value has to be bounded so that the scalar boson wave function is
quadratically integrable \cite{man,lane3}.
%%%%%%%%%%%%%%%%%%%%%%%%%%%%%%%%%%%%%%%%%%%%%%%%%%%%%%%%%%%%%%
\par The contribution due to the fermion loop indicated in Fig.(3a) , particularly in the
extreme walking behavior or massive top case for the self-energy (i.e. $\gamma \to 2({\rm or}\, \kappa \to 1)$),  tends to decrease $M^2_{\phi}$ according to Eq.(\ref{eq29}), and in this case  this contribution lowers the estimate of the composite scalar boson mass. 

\par In Fig.(3c), the vertex $\lambda$ can be approximately represented by
\be 
\lambda \approx \frac{g_{w}}{\pi}N_{F}\lambda_{ETC}\frac{\Lambda_{TC}}{M_W}\left(\frac{\Lambda^2_{TC}}{\Lambda^2_{ETC}}\right)^{1-\kappa},
\label{eq29b}
\ee 
\noindent where $N_F$ is the number of technifermions that couple to the fermions (f) in the loop,  and we assumed the existence of an ETC gauge theory with coupling $\alpha_{ETC}$. The effective charge , $\lambda_{ETC} = C_{ETC}\alpha_{ETC}$,   involve the ETC coupling  and the appropriate ETC Casimir operator eigenvalues $C_{ETC}$. The top quark makes the most significant contribution in the loop described in Fig.(3), as discussed in Refs.\cite{us00,us001} in the extreme walking behavior, its mass can approximately be expressed by $m_t(\gamma)\approx N_F\lambda_{ETC}\Lambda_{TC}$, so we can write  the vertex $\lambda$ as 

\be 
\lambda \approx \frac{g_{w}}{\pi}\frac{m_t(\gamma)}{M_W}\left(\frac{\Lambda^2_{TC}}{\Lambda^2_{ETC}}\right)^{1-\kappa}.
\label{eq29f}
\ee  

%%%%%%%%%%%%%%%%%%%%%%%%%%%%%%%%%%%%%%%%%%%%%%%%%%%%%%%%%%%%%%
\par In the limit when $\kappa \to 0$, we have 
$$
\delta M^2_{\phi}(0) \approx 0 ,
$$
\noindent while in the limit when $\kappa \to 1({\rm or}\, \gamma \to 2)$ , we recover the  effective  coupling of the top described in Eq.(\ref{eqmt})
$$
\lambda \propto  \frac{g_{w}}{\pi}\frac{m_t(\gamma)}{M_W}\approx \frac{m_t(\gamma)}{v}
$$
and  we obtain 
\br
&& \delta M^2_{\phi}(2) \approx -\frac{3\lambda^2 }{4\pi^2}\Lambda^2_{TC}\nonumber \\
&& \hspace*{1.3cm} \approx -\frac{3}{\pi^4}\left(\frac{m_t(2)}{v}\right)^2\Lambda^2_{TC}. 
\label{eqdelta}
\er
\par At this point we should highlight that in the parameterization of the estimate presented by Eq.(\ref{eqdelta}), $m_t(2)$ is model dependent. In Fig.(3), the number of technifermions $F$ ($N_Q$:techniquarks or $N_L$:technileptons) that generate the $m_t(2)$ mass depends on ETC interactions. However, we can consider as an illustrative example the model described in Ref.\cite{us001}, where in Fig.(3) of that reference, we present the diagrams that contribute to $m_t(2)$ in that work. 
As a result of these contributions $m_t(2)$ was estimated to be on the order of $\sim 100GeV$, what leads to
\be 
 |\delta M_{\phi}(2)|  \approx 70 GeV.
\ee 

\par As we have seen, the determination of $m_t(2)$ is model dependent,  however, assuming that it is possible to elaborate a more realistic ETC model , where in principle $m_t(2)$ can be of the same order of the observed top quark mass, the positivity condition of $M^2_S > 0$, which is given by the smallest 
BSE solution ($M_S(2)$) minus the fermionic correction described 
by Eq.(\ref{eqdelta}), leads to the following intriguing theoretical limit
\be 
\frac{M_S(2)}{\Lambda_{TC}} > \frac{\sqrt{3}}{\pi^2}\frac{m_t}{v} ,
\label{eqhm}
\ee 
i.e., assuming $\Lambda_{TC} \approx 1$TeV and use the known $m_t$ and $v$ values the bound of Eq.(\ref{eqhm}) is exactly of the order of the known Higgs boson mass,
or $M_S(2)\geq 123.3 GeV$. 

%%%%%%%%%%%%%%%%%%%%%%%%%%%%%%%%%%%%%%%%%%%%%%%%%%%%%%%%%%%%%%%

\par Note that these are very rough estimates originated by the existence of radiative corrections due to TC and ETC as appear in Fig.(3).
The effect of fermion loops inevitably decreases the scalar bound state mass.
A full BSE calculation should also involve the dependence of all Green's functions on the mass anomalous dimensions and fermion masses of
the scalar boson constituents. Actually, the result of this section can be seem only as a correction
to the BSE result of the previous section, which is dependent on the ansatz proposed in Eq.(\ref{eq4}), whereas a complete calculation
should rely on a self-energy obtained from the Schwinger-Dyson equation, which is beyond the scope of the present work. 
It is also opportune to recall that even the sigma meson
mass calculation may have corrections of similar type (generated by electroweak bosons exchange), that can lower the BSE evaluation of its mass.

\section{Conclusions}

In the case of a possible composite scalar boson, we computed its mass using Bethe-Salpeter equations 
and assuming constituents of same mass. The calculation was performed with the
help of a constituent self-energy dependent on the mass anomalous dimension. Our result indicates how the scalar masses, no matter we are talking about the 
sigma meson or the Higgs boson, can vary with the mass anomalous dimension as shown in Eqs.(\ref{eqg1}) and (\ref{eqg2}). We hope
that this behavior can be tested by other methods.

In Section III we call attention to the fact that a full BSE calculation should include diagrams like the one of Fig.(3). The effect
of such diagrams is to lower the scalar boson mass. As a simple estimate of this effect we have computed the fermionic contribution
of the radiative corrections induced by Fig.(3a). Of course, our calculation is very simple but it shows that this effect cannot be 
neglected. The bound of Eq.(\ref{eqhm}) is an example of the balance between the different contributions to scalar masses.

Our results using the Bethe-Salpeter equations show how scalar composite masses can be smaller than the composition scale, as long as we have large anomalous dimensions and the effect of
fermions, like the top quark, included into the calculation. Actually, we may have a delicate balance between the mass anomalous dimension 
of the fermionic constituents and the contribution of fermions that contribute negatively to the Higgs boson mass.

If the Higgs boson is a composite particle, it is still possible that its constituents are bounded by a non-Abelian gauge strong interaction
similar to QCD. In this case this new strong interaction dynamics can be scaled from the known QCD Green's functions, which nowadays are
used to describe hadronic physics with high accuracy as reported in Refs.\cite{cdr1,cdr2}.
Therefore, we believe that there is a systematic path to perform a realistic Higgs boson mass 
calculation using BSE and DSE, assuming a given TC gauge group with Green's functions scaled from QCD, and
varying data such as the number of fermions and other parameters until obtaining the Higgs boson mass experimental value.

\section*{Acknowledgments}
This research  was  partially supported by the Conselho Nacional de Desenvolvimento Cient\'{\i}fico e Tecnol\'ogico (CNPq)
under the grants 310015/2020-0 (A.D.) and 303588/2018-7 (A.A.N.) .

\end{document}